\documentstyle[aps,floats,prl,twocolumn,epsf]{revtex}

\begin{document} 
\draft

\preprint{cond-mat/9610168}

\date{Received by PRL: October 16, 1996} 

\title{Magnetization plateaus in spin chains: ``Haldane gap'' for
half-integer spins}
\author{Masaki Oshikawa$^1$, Masanori Yamanaka$^3$ and Ian Affleck$^{1,2}$}
\address{Department of Physics and Astronomy$^1$
and Canadian Institute for Advanced
Research$^2$, 
\\ University of British Columbia, Vancouver, BC, V6T 1Z1, CANADA
\\
Department of Applied Physics$^3$, University of Tokyo
Hongo, Bunkyo-ku, Tokyo 113 JAPAN
} 

\maketitle 

\begin{abstract}
We discuss zero-temperature quantum spin chains in
a uniform magnetic field, with axial symmetry.
For integer or half-integer spin, $S$,
the magnetization curve can have plateaus and
we argue that the magnetization per site $m$ is topologically
quantized as $n (S - m)= \mbox{integer}$ at the plateaus,
where $n$ is the period of the groundstate.
We also discuss conditions for the presence of the plateau
at those quantized values.
For $S=3/2$ and $m=1/2$, we study several models and find two
distinct types of massive phases at the plateau.
One of them is argued to be a ``Haldane gap phase''
for  half-integer $S$.
\end{abstract}


\narrowtext

One-dimensional antiferromagnets are expected not to have
long-range magnetic order in general.
It was argued by Haldane~\cite{Haldane:conj},
in 1983, that  for integer, but not half-integer spin,
$S$, there is a gap to the excited states.
In the presence of a magnetic field, 
the $S=1/2$ Heisenberg antiferromagnetic (AF) chain
remains gapless from zero field up to the saturation field,
where the groundstate is fully polarized~\cite{magcurve}.
For integer $S$, the gap persists up to a critical field, equal to the gap,
where bose condensation of magnons occurs~\cite{Affleck:bosecon}.
The $S=1$ Heisenberg AF chain is known to be gapless
from the critical field up to the saturation
field~\cite{SakaiTakahashi:S1mag}.
Recently Hida observed that an $S=1/2$ antiferromagnetic chain with
period $3$ exchange coupling shows a plateau in the magnetization curve
at magnetization per site
$m=1/6$ ($1/3$ of the full magnetization)~\cite{Hida:trim}.
Related works on bond-alternating chains have also
been reported~\cite{Okamoto:trim,Miya:trim,Tone:S1dim,Totsuka:dim}
including experimental observation~\cite{Hagi:dim}.

\begin{figure}[htbp]
\begin{center}
\leavevmode    
\epsfxsize=2.8in
\epsfbox{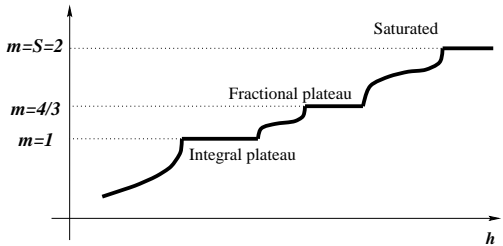}
\caption{Schematic diagram of possible magnetization curve
of an $S=2$ chain.
There are plateaus at special values of magnetization $m$.
The plateau with a fractional $S-m=2/3$ must accompany
a period $3$ groundstate.(See text for details.)}
\label{fig:schem}
\end{center}
\end{figure}

In this letter, we consider the zero-temperature behaviour of
general quantum spin  chains, including chains with periodic structures,
in a uniform magnetic field pointing along the direction of
the axial symmetry ($z$-axis). (i.e. the total $S^z$ is conserved.)
We argue that, in quantum spin chains, there is a phenomenon
which is strikingly analogous to the Quantum Hall Effect
-- topological quantization of a physical quantity under
a changing magnetic field~\cite{QHE}.(See Fig.~\ref{fig:schem}.)
We first consider an extension of the Lieb-Schultz-Mattis (LSM)
theorem~\cite{LSM} to the case with an applied field.  
This indicates that translationally invariant spin chains in an
applied field can be gapful without breaking translation symmetry,
{\em only when} the magnetization per spin, $m$, obeys $S-m =$ integer.
We expect such gapped phases to correspond to plateaus at
these quantized values of $m$.
``Fractional quantization'' can also occur,
if accompanied by (explicit or spontaneous)
breaking of the translational symmetry.
The generalized LSM theorem does not prove
the presence of the plateau, however.
Thus we construct a corresponding argument using Abelian bosonization,
which is in complete agreement with the generalized LSM theorem,
and also gives a condition for the presence of the plateau.
As simplest examples, we study translationally invariant
$S=3/2$ chains at $m=1/2$.
We present numerical diagonalization and
Density Matrix Renormalization Group (DMRG)~\cite{White:DMRG}
calculations, which demonstrate the existence of the two distinct types
of gapped phases for generalized models.
They are related to the $S=1$ large-$D$ phase and the $S=1$ Haldane phase,
respectively.
On the other hand, our study shows that
the standard $S=3/2$ Heisenberg model is gapless with no plateau at $m=1/2$.
We only give a brief summary of our numerical work in this letter;
details and further results, including the effect of
the axial symmetry breaking, will be presented
in a longer paper~\cite{inprep}.

The LSM theorem~\cite{LSM} proves
the existence of at least one low-energy,
$O(1/L)$ excited state for even length $L$ half-integer $S$
AF chains with periodic boundary conditions and general, translationally
invariant Hamiltonians.
It is expected that, this implies
either gapless excitations or spontaneously broken translational symmetry
in the $L\to \infty$ limit.  The failure of this proof for the
integer-$S$ case is necessary for the existence of the Haldane phase
with no broken translational symmetry and a gap. 
We observe that the original version
of this proof also works in a magnetic field
{\it except for integer $S-m$}.
Thus only in this case is a massive phase without spontaneously
broken translational symmetry possible.
The proof consists of making a slow rotation on the groundstate, 
$|\psi \rangle$, assumed to be unique,
and observing that the resulting low-energy state is orthogonal
to the groundstate.  The rotation operator is
$U \equiv \exp{[- i\sum_{j=1}^L(2\pi j/L)S^z_j]}$.
For any Hamiltonian $H$, including a magnetic field term,
with short-range interactions which is invariant
under rotation about $z$-axis,
it can be shown that
\begin{equation} 
\langle \psi |U^\dagger HU-H|\psi \rangle= O ( \frac{1}{L} ).
\end{equation} 
This implies the existence of an excited state
with excitation energy of $O(1/L)$,
if we can show that $U|\psi \rangle$ is orthogonal to $|\psi \rangle$. 
To this end, we use the invariance of $H$ under translation by one site: $T$.
This operation maps $U$ into:
\begin{equation} 
U \to T U T^{-1} = Ue^{i2\pi S^z_1 - i(2\pi/L) \sum_{j=1}^LS^z_j}.
\end{equation}
Namely, the operation of $U$ changes the eigenvalue
of $T$ by a factor $e^{i 2 \pi (S-m)}$, 
where $m = \sum_{j=1}^L S^z_j / L$.
Thus $U|\psi \rangle$ must be orthogonal to $|\psi \rangle$ except when
$(S-m)$ is an integer. 
We note that this is consistent with previous
results for translationally invariant $S=1/2$ and $1$ AF chains,
where no gap is found at partial magnetization.
However for higher spin, gapped phases at partial magnetization
are possible without breaking the translational symmetry,
when $S-m$ is an integer.

When $S-m$ is not an integer, there is a low-lying state with
energy of $O(1/L)$.
This means either a massless phase with a continuum of low-energy
states or spontaneous symmetry breaking in the thermodynamic limit.
Following the above proof,
when $S-m = p/q$ where $p$ and $q$ are coprimes,
$U^k | \psi \rangle$ for $k=0, 1, \ldots, q-1$ have different eigenvalues
of $T$. Thus these $q$ states have low energy of $O(1/L)$.
If these are related to a spontaneous breaking of the symmetry,
the ground states in the thermodynamic limit should be $q$-fold
degenerate. Since they have $q$ different eigenvalues of $T$,
they can be related to a spontaneous breaking of
the translation symmetry to period of $q$ sites
in the thermodynamic limit.
It is natural to expect a gap and plateau in this case.
As in the case of Quantum Hall Effect,
``fractional quantization'' is therefore possible, accompanying
the spontaneous breaking of the translation symmetry
in the present case.(See Fig.~\ref{fig:schem})
We may compare this to a hidden symmetry breaking
in Fractional Quantum Hall Effect~\cite{TaoWu:FQHE}.

Possibly there is a low-energy state other than constructed as above.
In such a case, we can again construct a set of $q$ low-energy states
by operation of $U$. Thus in general the period of ground state can
be an integral multiple of $q$.
A simple example of such a case is known for $S=1$ and $m=0$.
While a gap without spontaneous symmetry breaking is possible as in
the Haldane phase, spontaneous dimerization
occurs in the bilinear-biquadratic Heisenberg model
for a range of parameters~\cite{AKLT}.

Our generalization of LSM theorem is easily extended to
Hamiltonian with spatial structures:
bond-alternating chains~\cite{Hida:trim,Tone:S1dim}, 
spin-alternating chains~\cite{Pati:alt}, spin ladders, etc.
For example, Hida's model~\cite{Hida:trim} is only invariant
under a three-site translation $T^3$; a massive phase without
spontaneous symmetry breaking is possible for $3(S-m) = \mbox{integer}$.
Thus a quantized plateau is possible at $m=1/6$ as he observed.
In general, the quantization condition is given by
$ S_u - m_u = \mbox{integer}$, where $S_u$ and $m_u$ are
respectively the sum of $S$ and $m$ over all sites in the
unit period of the groundstate.
The period of the groundstate is determined by
the explicit spatial structure of the Hamiltonian,
and also by spontaneous symmetry breakings.

The low-energy state $U | \psi \rangle$ appearing in the
LSM theorem has the same total magnetization as
in the groundstate.
It does not directly contradict the existence of a plateau,
which is determined by the gap to states with other
total magnetizations.
However, we expect that, in general a gapless phase has
low-energy states in both fixed and different magnetization sector,
as can be seen in the following Abelian bosonization approach.
Schulz~\cite{Schulz86} explained the difference
between integer and half-integer
spin by Abelian bosonization.
We show that his result can be understood more simply as
a consequence of symmetries.
At the same time, we generalize the discussion to the case with
a non-vanishing magnetization.
Following Schulz~\cite{Schulz86}, we 
start from Abelian bosonization of $2S$ spin-$1/2$ chains
and then couple them to form a spin-$S$ chain.
Firstly, each spin-$1/2$ chain is fermionized by Jordan-Wigner
transformation.
The $z$ component of each spin-$1/2$ is related to the fermion
number as $ \sigma^z_n = 1 - 2 \psi^{\dagger}_n \psi_n $.
Then the low-energy excitations are treated
by  continuous fermion fields.
Let us denote the lattice spacing as $a$ and
the spatial location $x = na$. 
The continuous fermion fields $\psi_R$ and $\psi_L$ are
defined by
\begin{equation}
  \psi^j_n \sim e^{i k_F x} \psi^j_R(x) + e^{-i k_F x} \psi^j_L(x),
\label{eq:contfermi}
\end{equation}
where $j = 1 , \cdots, 2S$ is the ``flavor'' index to distinguish
$2S$ spin-$1/2$'s.
They are bosonized in a standard way:
$  \psi^j_{R} = e^{ i \varphi^j_{R}/R} $ and
$  \psi^j_{L} = e^{ - i \varphi^j_{L}/R} $,
where $\varphi^j_R$ and $\varphi^j_L$ are chiral bosons and
$R$ is the compactification radius of the boson.
$R$ will be renormalized by interactions~\cite{Affleck:LesHouches},
and will eventually depend on the model and on the magnetization $m$.
(For an isotropic model, $R$ is fixed by the symmetry at $m=0$, but
the magnetic field breaks the symmetry and thus $R$ will depend on $m$.)
We define the non-chiral bosonic field
$  \varphi^j = \varphi^j_L + \varphi^j_R $ and its dual
$  \tilde{\varphi}^j = \varphi^j_L - \varphi^j_R$ .

Interactions among bosonic fields are also generated during
the mapping from the original spin problem.
In general, we expect any interaction would be generated if not
forbidden by a symmetry.
Thus we analyze symmetries of the system, following the treatment
of spin-$1/2$ chains in Ref.~\cite{Affleck:LesHouches}.
The original problem has a $U(1)$ symmetry: rotational invariance
about $z$ axis.
Rotation of each spin-$1/2$ is given by the phase transformation
$  \psi^j_{L,R} \rightarrow e^{i \theta} \psi^j_{L,R}$
of the corresponding fermion.
In bosonic language, this corresponds to a shift of the dual field
$  \tilde{\varphi}^j \rightarrow \tilde{\varphi}^j + \mbox{const.} $.
Since we have coupled $2S$ spin-$1/2$ chains into a spin-$S$ chain,
only the simultaneous rotation of $2S$ spin-$1/2$'s is a symmetry
of the system.
If we define a new bosonic field $\phi = \sum_j \varphi^j$
(and similarly for $\tilde{\varphi}^j$),
the $U(1)$ symmetry is written as
$  \tilde{\phi} \rightarrow \tilde{\phi} + \mbox{const} $.
Thus all the interactions of the form $e^{\pm 2 n \pi i R \tilde{\phi}}$
are prohibited by the symmetry.
The remaining $2S-1$ fields, which are defined by linear combinations
of original $\tilde{\varphi}^j$ fields, are not protected by the symmetry.
Thus all fields except $\phi$ are expected to become
massive by interactions, as Schulz observed by an explicit calculation.
The remaining $\phi$ field, is also subject to $e^{\pm i n \phi/R}$
type interactions.
Let us consider another symmetry of the system: one-site translation.
By definition~(\ref{eq:contfermi}),
it actually corresponds to a transformation of the continuum field
$  \psi^j_R  \rightarrow  e^{i k_F a} \psi^j_R $ and
$ \psi^j_L  \rightarrow  e^{- i k_F a} \psi^j_L $.
Again, only the simultaneous translation of all flavors is a symmetry
of the system. 
Thus the one-site translation $T$ is written as
$  \phi \rightarrow \phi + 4 S (k_F a) \pi R $,
in the bosonic language.

Since all of the $2S$ flavors are equivalent, the magnetization
should be equally distributed among them.
Thus the fermi momentum $k_F$ is determined as 
$  k_F a = (S - m) \pi / (2S) $ .
As a consequence, the one-site translation $T$  is given by
\begin{equation}
  \phi \rightarrow \phi + 2 (S - m) \pi R .
\end{equation}
Thus the leading operator $\cos{(\phi/R)}$ is permitted only if
$S - m$ is an integer.
For $m$ satisfying the quantization condition,
the leading operator $\cos{(\phi/R)}$
should be relevant in order to produce a gap.
Thus $R$ must be larger than $R_c = 1/\sqrt{ 8 \pi}$
for the presence of the plateau.
If $S - m = p/q$ where $p$ and $q$ are coprimes,
the operator $\cos{( q \phi /R) }$ is permitted.
It can be relevant if $R \geq q / \sqrt{ 8 \pi}$
(this is a severe condition for a large $q$);
if it is, a groundstate in the thermodynamic limit
corresponds to a potential minimum of $\cos{( q \phi /R)}$.
There are $q$ such groundstates and they are mapped to each
other by applying the translation operator $T^k$ ($k < q$).
Thus the ground states have spontaneously $q$-fold broken
translation symmetry.
These results are in agreement with the generalized LSM theorem,
and also give conditions for a finite plateau at the quantized values.

Our bosonization argument is also readily generalized to
models with spatial structures.
Our picture is consistent with Okamoto's
analysis~\cite{Okamoto:trim} of Hida's plateau~\cite{Hida:trim}.
For  $S=1$ AF chains,
Tonegawa et al.~\cite{Tone:S1dim}
obtained an $m=1/2$ plateau as soon as they introduced 
a small bond-alternation. In our approach, the leading operator
is expected to appear as soon as the translational
symmetry is broken. 
Thus we expect a plateau for any finite amount of bond-alternation,
if the radius exceeds the critical value, in agreement
with Ref.~\cite{Tone:S1dim}.
This is also in agreement with an explicit bosonization calculation by
Totsuka~\cite{Totsuka:dim} for $S=1$ bond-alternating chains.

Now let us discuss some examples of translationally invariant
$S=3/2$ chains.
It is interesting both from an experimental and conceptual point of view
to add an easy-plane crystal field term:
\begin{equation}
  H = \sum_j \vec{S}_j \cdot \vec{S}_{j+1} + D (S^z_j)^2 - h S^z_j .
\label{eq:Dham}
\end{equation}
Clearly, if $D>>1$, all the spins are first fixed to $S^z=1/2$
with increasing field before any of the spins go into the $S^z=3/2$ state,
corresponding to a gapped $m=1/2$ plateau.
The presence of finite gap and plateau is
proved rigorously for a sufficiently large but
finite $D$~\cite{Hal:priv}, by applying the general theorem
in Ref.~\cite{KennedyTasaki:hid}.
This situation is reminiscent of that which occurs in
the large-$D$ phase in a zero field $S=1$ chain. 
Numerically, we found
a finite $m=1/2$ plateau at least for $D \geq 2$.

\begin{figure}
\begin{center}
\leavevmode    
\epsfxsize=2.6in
\epsfbox{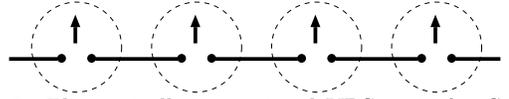}

\caption{The partially magnetized VBS state for $S=3/2$.
A solid line denotes a valence bond (singlet formed from
two spin-$1/2$'s). 
An up arrow denotes a spin-$1/2$ with $S^z=1/2$.
A dashed circle represents the symmetrization of
spin-$1/2$ variables at each site.}
\label{fig:pmvbs}
\end{center}
\end{figure}

Another kind of
trial groundstate for an $S=3/2$ chain, corresponding
to an $m=1/2$ plateau is shown in Fig.~\ref{fig:pmvbs}
in the valence bond notation~\cite{AKLT}.
Regarding each $S=3/2$ operator as being a symmetrized
product of three $S=1/2$'s, one
$S=1/2$ is polarized by the applied field at each site while the other
two form a valence-bond-solid (VBS)
groundstate as occurs for an $S=1$ chain in zero field.
In Ref.~\cite{MOVBS}, this partially magnetized VBS state was proposed
and the relevance to the magnetization process was suggested.
A generalization of this kind of VBS-type state and
further analysis were later done in Ref.~\cite{Niggemann}.
(See also~\cite{Pati:alt}.)
Clearly this sort of VBS-type state
exists for all $S$ and $m$ such that $S - m$ is an integer.

We can construct a model to realize the $S=3/2$ VBS-type
state in Fig.~\ref{fig:pmvbs} as a ground state:
\begin{equation}
  H = \sum_j  P^{(j,j+1)}_3 + \alpha \vec{S}_j \cdot \vec{S}_{j+1} 
                - h S^z_j ,
\label{eq:P3ham}
\end{equation}
where $P^{(j,j+1)}_3$ is the projection operator onto the space  
with total spin $3$ for sites $j$ and $j+1$.
At $\alpha = 0$, any state constructed
with one valence bond between neighboring sites is a groundstate.
The ground state is thus infinitely degenerate due to the ``free''
spin-$1/2$ at each site.
Applying an infinitesimal magnetic field, the degeneracy is lifted and the
ground state is the above mentioned VBS-type state (Fig.~\ref{fig:pmvbs}).
Thus the model with $\alpha = 0$ has an $m=1/2$ plateau starting from
zero magnetic field.
Turning on the Heisenberg term, $\alpha$, the degeneracy at $h=0$
is lifted and a finite magnetic field is required to reach $m=1/2$.
For small value of $\alpha$, however, we might still expect a
finite $m=1/2$ plateau.
We studied this model with periodic boundary conditions
by numerical diagonalization for up to $12$ sites
and found the $m=1/2$ plateau exists at least for $\alpha \leq 0.06 $.
In contrast to the plateau at large positive $D$, which
is related to the large-$D$ phase in $S=1$ chains,
it is natural to relate this state to the $S=1$ Haldane phase.

For $S=1$, the Haldane phase is known to be distinct from the large-$D$
phase; these two massive phases are separated by a critical point
$D_c$, where the gap vanishes~\cite{Schulz86,BotetJulienKolb}.
The Haldane phase is characterized by the existence of
a topological long-range order~\cite{stringorder},
and gapless edge excitations in the open boundary
conditions~\cite{Kennedy:edge}.
These are understood as consequences of
a hidden symmetry breaking~\cite{KennedyTasaki:hid}.
One might suspect that the two
types of $S=3/2$ massive phases at the $m=1/2$ plateaus
discussed above, correspond to distinct phases.

If they are distinct,
there should be a phase transition between them.
In terms of Abelian bosonization, this phase transition may be
understood as the vanishing of the coefficient of the allowed
relevant operator $\cos{(\phi/R)}$, as in the case of $S=1$~\cite{Schulz86}.
We numerically measured the gap (width of the plateau) for the model:
\begin{equation}
  H =  \sum_j \alpha \vec{S}_j \cdot \vec{S}_{j+1} + D(S^z_j)^2
             +  P^{(j,j+1)}_3 - h S^z_j,
\label{eq:interpham}
\end{equation}
which interpolates between~(\ref{eq:Dham}) and~(\ref{eq:P3ham}).
For $\alpha=0.03 $ (fixed), we find the plateau vanishes
at $D \sim 4.5$, separating
the ``Haldane gap'' type plateau and the ``large-$D$'' type plateau.
Moreover, we compared the spectrum at $\alpha=0.03$ and $D=0$
between open and periodic boundary conditions, and found  evidence
for edge states. In the large-$D$ region, there are no such edge states.
These indicate that the ``Haldane phase'' at $m=1/2$ plateau,
which accompanies the edge states, 
is distinct from the ``large-$D$ phase''.

We also numerically examined the standard $S=3/2$ Heisenberg
AF chain with open boundary conditions,
by DMRG up to $100$ sites.
We did not find an $m=1/2$ plateau in this case,
in agreement with Refs.~\cite{Hida:trim,Okamoto:trim,Miya:trim}.
We emphasize that the absence is not a priori obvious.
As we have shown, in terms of the free boson theory, 
the plateau would be present if the compactification radius $R$
is greater than the critical value $R_c$ and the coefficient of
the most relevant operator $\cos{(\phi/R)}$ is non-vanishing.
We have determined the compactification radius from the spectrum 
for the open boundary condition obtained by DMRG,
as $R = 0.95 R_c < R_c$.
We note that $R$ is rather close to the critical value, and
possibly we can realize a plateau by a small perturbation of the
standard Heisenberg Hamiltonian~\cite{inprep}.
On the other hand,
while the radius is not completely well-defined 
at the {\em massive} $D=2$ plateau, a similar analysis gives the estimate
$R \sim 1.2 R_c > R_c$.
This result is consistent with the presence of the plateau.

The plateaus that we have found are closely related~\cite{inprep}
to Mott insulating (or charge density wave) phases in models of
interacting fermions or
bosons~\cite{Giamarchi:Mott1D,Niyaz:boseHubbard}.
Similarly to those cases~\cite{Schulz:IC-C,inprep}, we have found that
the singular part of the magnetization curve near a plateau
is proportional to $\sqrt{|h-h_c|}$
where $h_c$ is the critical field at (either) edge of a plateau,
at least for examples we have studied.
Our approach will also give new insights into
models of interacting fermions or bosons~\cite{inprep}.

It is a pleasure to thank Hal Tasaki for 
his collaboration in the early stage of this work and
many stimulating discussions.
We also thank Y. Narumi, S. Sachdev and T. Tonegawa for useful
correspondences.
The numerical work was partly based on program packages KOBEPACK/1 by
T. Tonegawa et al. and TITPACK ver. 2 by H. Nishimori.
The computation in this work has been done partially at
Supercomputer Center, ISSP, Univ. of Tokyo.
This work is partly supported by NSERC.
M.O. and M.Y. are supported by UBC Killam Memorial Fellowship and by JSPS
Research Fellowships for Young Scientists,
respectively.

{\bf Note added (January 1997):}
After the submission of the present letter,
we received a preprint by Totsuka,
which is a substantial enhancement
of his presentation~\cite{Totsuka:dim}
at Japanese Physical Society meeting Fall 1996,
and contains some of our general argument using bosonization.


\clearpage

\begin{thebibliography}{10}

\bibitem{Haldane:conj}
F.~D.~M. Haldane, Phys. Lett. {\bf 93A},  464  (1983).

\bibitem{magcurve}
R.~B. Griffiths, Phys. Rev. {\bf 133},  A768  (1964).

\bibitem{Affleck:bosecon}
I. Affleck, Phys. Rev. B {\bf 43},  3215  (1991).

\bibitem{SakaiTakahashi:S1mag}
T. Sakai and M. Takahashi, Phys. Rev. B {\bf 43},  13383  (1991).

\bibitem{Hida:trim}
K. Hida, J. Phys. Soc. Jpn. {\bf 63},  2359  (1994).

\bibitem{Okamoto:trim}
K. Okamoto, Solid State Commun. {\bf 98},  245  (1996).

\bibitem{Miya:trim}
M. Roji and S. Miyashita, J. Phys. Soc. Jpn. {\bf 65},  1994  (1996).

\bibitem{Tone:S1dim}
T. Tonegawa, T. Nakao, and M. Kaburagi,
J. Phys. Soc. Jpn. {\bf 65}, 3317 (1996).

\bibitem{Totsuka:dim}
K. Totsuka, JPS meeting Fall 1996, at Yamaguchi Univ.

\bibitem{Pati:alt}
S.~K. Pati, S. Ramasesha, and D. Sen, preprint {\tt cond-mat/9610080};
A.~K. Kolezhuk, H.-J. Mikeska, and S. Yamamoto, preprint
{\tt cond-mat/0610097}.

\bibitem{Hagi:dim}
Y. Narumi, K. Kindo, and M. Hagiwara, JPS meeting Fall 1996 at
Yamaguchi Univ., and private communications.

\bibitem{QHE}
R.~B. Laughlin, Phys. Rev. B {\bf 23},  5632  (1981);
D.~J. Thouless, M. Kohmoto, M.~P. Nightingale, and M. den Nijs, Phys. Rev.
 Lett. {\bf 49},  405  (1982).

\bibitem{LSM}
E.~H. Lieb, T. Schultz, and D.~J. Mattis,
Ann. Phys. (N.Y.) {\bf 16},  407 (1961);
I. Affleck and E.~H. Lieb, Lett. Math. Phys. {\bf 12},  57  (1986).

\bibitem{White:DMRG}
S.~R. White, Phys. Rev. B {\bf 48},  10345  (1993).

\bibitem{inprep}
M. Oshikawa, M. Yamanaka, and I. Affleck, in preparation.

\bibitem{TaoWu:FQHE}
R. Tao and Y.-S. Wu, Phys. Rev. B {\bf 30},  1097  (1984).

\bibitem{Schulz86}
H.~J. Schulz, Phys. Rev. B {\bf 34},  6372  (1986).

\bibitem{Affleck:LesHouches}
I. Affleck,  in {\em Fields, Strings and Critical Phenomena}, {\em Les Houches,
  Session XLIX}, edited by E. Brezin and J. Zinn-Justin (North-Holland,
  Amsterdam, 1988).

\bibitem{Hal:priv}
H. Tasaki, private communications.

\bibitem{KennedyTasaki:hid}
T. Kennedy and H. Tasaki, Commun. Math. Phys. {\bf 147},  431  (1992).

\bibitem{AKLT}
I. Affleck, T. Kennedy, E. Lieb, and H. Tasaki, Commun. Math. Phys. {\bf 115},
  477  (1988).

\bibitem{MOVBS}
M. Oshikawa, J. Phys. Condens. Matter {\bf 4},  7469  (1992).

\bibitem{Niggemann}
H. Niggemann and J. Zittartz, Z. Phys. B {\bf 101}, 289 (1996).

\bibitem{BotetJulienKolb}
R. Botet, R. Julien and M. Kolb, Phys. Rev. B {\bf 28}, 3914 (1983).

\bibitem{stringorder}
M. den Nijs and K. Rommelse, Phys. Rev. B {\bf 40},  4709  (1989);
S.~M. Girvin and D.~P. Arovas, Phys. Scr. T {\bf 27},  156  (1989);
Y. Hatsugai and M. Kohmoto, Phys. Rev. B {\bf 44},  11789  (1991).

\bibitem{Kennedy:edge}
T. Kennedy, J. Phys. Condens. Matter {\bf 2},  5737  (1990).

\bibitem{Giamarchi:Mott1D}
T. Giamarchi, to be published in the proceedings of the SCES96 conference
 (cond-mat/9609114) and references therein.

\bibitem{Niyaz:boseHubbard}
P. Niyaz, R.~T. Scalettar, C.~Y. Fong and G.~G. Batrouni,
Phys. Rev. B {\bf 50}, 362 (1994) and references therein.

\bibitem{Schulz:IC-C}
H.~J. Schulz, Phys. Rev. B {\bf 22} 5274 (1980).

\end{thebibliography}
\end{document}